\documentclass[reprint,amsmath,amssymb,twocolumn]{revtex4-2}
\usepackage[colorlinks,citecolor=blue,linkcolor=blue,anchorcolor=blue,filecolor=blue,
urlcolor=blue]{hyperref}
\usepackage{graphicx}
\usepackage{ulem}
\usepackage{dcolumn}
\usepackage{bm}
\usepackage{todonotes}

\usepackage{nccmath}

\usepackage{xfrac}

\usepackage{multirow}

\newcommand{\sat}{\mathrm{sat}}
\newcommand{\sym}{\mathrm{sym}}

\newcommand{\jel}{\mbox{\tiny{JEL}}}

\newcommand{\corr}{\mathrm{corr}}
\newcommand{\pot}{\mathrm{pot}}
\newcommand{\kin}{\mathrm{kin}}
\newcommand{\mass}{\mathrm{mass}}
\newcommand{\sky}{\mathrm{sky}}
\newcommand{\mm}{\mathrm{mm}}
\newcommand{\NL}{\mathrm{NL}}
\newcommand{\DD}{\mathrm{DD}}

\newcommand{\intern}{\mathrm{int}}

\usepackage{hyperref}

\begin{document}

\title{Nucleonic models at finite temperature with in-medium effective fields}


\author{Odilon Louren\c{c}o$^{1,2}$, Mariana Dutra$^{1,2}$, J\'er\^ome Margueron$^{1,3}$}

\affiliation{
$^1$Institut de Physique des 2 infinis de Lyon, CNRS/IN2P3, Universit\'e de Lyon, Universit\'e Claude Bernard Lyon 1, F-69622 Villeurbanne Cedex, France 
\\
$^2$Departamento de F\'isica, Instituto Tecnol\'ogico de Aeron\'autica, DCTA, 12228-900, S\~ao Jos\'e dos Campos, SP, Brazil 
\\
$^3$International Research Laboratory on Nuclear Physics and Astrophysics, Michigan State University and CNRS, East Lansing, MI 48824, USA
}

\date{\today}

\begin{abstract}
We perform a calculation of dense and hot nuclear matter where the mean interaction between nucleons is described by in-medium effective fields and where we employ analytical approximations of the Fermi integrals. We generalize a previous work, \cite{dutra2023finite}, where we have addressed the case of the Fermi gas model with in-medium effective mass. In the present work, we fully treat the in-medium interaction by considering both its contribution to the in-medium effective fields, which can be subsumed by the mass in some cases, and to the potential term. Our formalism is general and could be applied to relativistic and nonrelativistic approaches. It is illustrated for different popular models -- Skyrme, nonlinear, and density-dependent relativistic mean-field models --, but also for the metamodel, and it provides a clear understanding of the in-medium correction to the pressure, which is present in the case of the Skyrme and metamodel but is not for the relativistic ones. For the Fermi integrals, we compare the analytical approximation to the, so-called, ``exact'' numerical calculations in order to quantitatively estimate the accuracy of the approximation.
\end{abstract}

\maketitle

\section{Introduction}

The description of dense matter depends to a large extend on the nuclear interaction, which is expressed within various models~\citep{stone2007,baoanli2008,batista,gogelein,dutra2012,dutra2014}. In the present work, we consider phenomenological approaches for the nuclear interaction, for which we suggest a common formalism at finite temperature. In addition, at variance with zero temperature where the Fermi integrals are analytical, nuclear matter at finite temperature often requires the numerical calculation of Fermi integrals, which represents a numerical cost and impacts the computing time. In particular, the use of statistical methods such as the Bayesian statistics coupled to Markov chain Monte-Carlo, which are more and more employed to accurately quantify uncertainties, requires a large number of calculations. In this case, it is crucial to reduce all possible sources of extra-time consumption at finite temperature and, for instance, to employ analytical approximations of the Fermi integrals. 

In a previous work, \cite{dutra2023finite}, we have shown how in-medium corrections to the nucleon effective mass could be incorporated into the Fermi gas model (FG), which is a generalization of the free Fermi gas one (FFG). Phenomenological models for nucleon interaction at the mean-field approximation predict indeed in-medium correction to the effective mass, and more generally in-medium modification of effective fields, which was not treated in our previous work. These fields could be the in-medium effective mass or the in-medium meson fields, or any other fields which induce an implicit medium correction to thermodynamical quantities. In the present paper, we treat entirely the interaction term at the mean-field level provided the in-medium corrections could be devised into an in-medium effective mass and a momentum independent mean-field terms. Our formalism is however limited to models where the momentum dependence of the interaction can be represented by a modification of the bare mass, such as in Skyrme,  relativistic mean field, and metamodel approaches. Given this limitation, we present a formalism where the full contribution of the interaction is considered at finite density and temperature, making use of a fast analytical approximation of the Fermi integrals.

The formalism which is shown in this paper is directly employable to perform finite temperature calculations based on phenomenological nucleon interaction. We compute finite temperature calculation for dense matter based on the time consuming, but ``exact'', calculation of the Fermi integrals, which is compared to its analytical approximation using the one suggested in \cite{JEL-1996-1020}, hereafter called JEL. The suggested formalism allows one to perform fast calculations at finite temperature, and in dense and uniform matter existing in the dense phases of core-collapse supernovae~\citep{bethe,Sumiyoshi_2019} or in the remnants of neutron star mergers~\citep{shibata:2019}.


Our work is organized as follows: In Sec.~\ref{sec:formalism} we perform the generalization of the FG model described in~\cite{dutra2023finite} by introducing, in the canonical ensemble, the Helmholtz free energy density for the full interaction term including in-medium fields for relativistic and nonrelativistic models. In Sec.~\ref{sec:examples} we apply this formalism to Skyrme, nonlinear, density-dependent relativistic mean-field, and metamodel, and generate thermodynamical quantities, such as the pressure and the chemical potential, with in-medium corrections induced by the fields. Finally, our conclusions are presented in Sec.~\ref{sec:conclusions}.

\section{Thermodynamical description of hot and dense matter}
\label{sec:formalism}

In the following, we consider the canonical ensemble (CE) allowing exchanges of energy in open nuclear systems controlled on average by the temperature (intensive variable), but with a fixed number of particles (extensive variables). For a system composed of neutrons and protons, these densities are $n_n$ and $n_p$. Equivalently, one could describe this system with the nucleonic density $n=n_n+n_p$ and isospin parameter $\delta=(n_n-n_p)/n$. Due to the equivalence between ensembles in infinite and uniform systems, one could replace particle numbers by chemical potentials, which control the average number of particles in the Grand-Canonical ensemble. In the following, the CE will however be adopted since it is more frequent to employ densities instead of chemical potentials.

In the CE,  the thermodynamical potential is defined to be the Helmholtz free-energy density $\phi$, which is expressed in terms of the energy density $\epsilon$ and entropy density $\sigma$ as
\begin{align}
\phi &\equiv \epsilon - T\sigma\, . 
\label{eq:Helmholtz}
\end{align}
It can be decomposed into a kinetic and a potential contributions as,
\begin{align}
&\phi(n,\delta,T,\{\varphi\}) = \phi^*_{\kin}(n,\delta,T,\{\varphi\})
+ \phi_{\pot}(n,\delta,\{\varphi\})\, , \label{eq:phi}
\end{align}
where $\{\varphi\}$ stands for a set of field contributions $\varphi_{i}$, which could depend on the thermodynamical variables $n$, $\delta$ or $T$. 
The detailed expression of $\{\varphi\}$ depends on the model for which this formalism is applied to. The general notation adopted in this section is illustrated in the next section. For instance, we could have $\{\varphi\}=\{m^*_n,m^*_p\}$ in the case of the Skyrme interaction, while in relativistic mean field approaches, the fields are those of the meson contributions to the mean-field. 

The kinetic term originating from the neutron and proton contributions can be expressed as, with $q$ representing neutrons or protons,
\begin{align}
\phi^*_{\kin}(n,\delta,T,\{\varphi\}) &= \sum_{q=n,p}
\phi^*_{\kin,q}(n,\delta,T,\{\varphi\}_q)\, ,
\end{align}
where the two kinetic terms $\phi^*_{\kin,q}$ are those of the FG corrected by a field $\{\varphi\}_q=m^*_q$, the density-dependent nucleon effective mass, see Ref.~\cite{dutra2023finite} for more details. We consider the notation introduced in Ref.~\cite{dutra2023finite} where the thermodynamical quantities with $^*$, like $\phi^*$ for instance, are calculated using analytical expressions valid at fixed and constant in-medium effective mass. Note that the mass is not necessarily taken to be the bare mass, but the correction due to its variation as a function of the thermodynamical variables is not incorporated in quantities with $^*$. In other words, the quantities with $^*$ are the ones that are calculated directly using analytical expressions, such as the ones given in the JEL approximation. As noted in Ref.~\cite{dutra2023finite} some thermodynamical properties calculated by using the JEL approximation, for instance, shall be corrected by the modification of the in-medium effective mass, which is not given by the JEL approximation.

In eq.~\eqref{eq:phi}, the term $\phi_{\pot}(n,\delta,\{\varphi\})$ is the potential contribution, which is considered as (explicitly) independent of $T$ in the present work: In general, phenomenological nucleonic potentials do not explicitly depend on the temperature. 

The pressure of the system is obtained from the Helmholtz free energy per particle, $f(n,\delta,T,\{\varphi\}) = \phi(n,\delta,T,\{\varphi\})/n$, as
\begin{align}
p &= n^2 \frac{\partial f}{\partial n}\Bigg|_{T,\delta} = n^2\frac{\partial f^*_{\kin}}{\partial n}\Bigg|_{T,\delta,\{\varphi\}}
\nonumber\\ 
&+ n^2\sum_{i}
\frac{\partial \varphi_{i}}{\partial n}\Bigg|_{T,\delta}
\frac{\partial f}{\partial \varphi_{i}}\Bigg|_{T,n,\delta,\{\varphi_{{j\neq i}}\}}
\hspace{-0.9cm}+ n^2\frac{\partial f_{\pot}}{\partial n}\Bigg|_{T,\delta,\{\varphi\}}
\nonumber\\
& = \sum_{q=n,p} p^*_{\kin,q} \,+ p_{\corr} + p_{\pot}\, ,
\label{eq:press}
\end{align}
with $p^*_{\kin,q}=n^2 \partial f^*_{\kin,q}/\partial n\vert_{T,\delta,\{\varphi\}}$ and $p_\corr$ the correction due to the implicit density dependency of the fields, 
\begin{align}
p_\corr &= n^2\sum_i
\frac{\partial \varphi_{i}}{\partial n}\Bigg|_{T,\delta}
\frac{\partial f}{\partial \varphi_{i}}\Bigg|_{T,n,\delta,\{\varphi_{{j\neq i}}\}} 
\nonumber\\
&= n\sum_i
\frac{\partial \varphi_{i}}{\partial n}\Bigg|_{T,\delta}
\frac{\partial \phi}{\partial \varphi_{i}}\Bigg|_{T,n,\delta,\{\varphi_{{j\neq i}}\}} \, 
\label{eq:pcorr}
\end{align}
where we employ the usual notation: for the particle number $i$, ${{j\neq i}}$ means all other particle numbers. The potential contribution to the pressure is defined as,
\begin{align}
p_{\pot} = 
n^2\frac{\partial f_{\pot}}{\partial n}\Bigg|_{T,\delta,\{\varphi\}}
= n\frac{\partial\phi_{\pot}}{\partial n}\Bigg|_{T,\delta,\{\varphi\}} - \phi_{\pot} \, .
\label{eq:ppot}
\end{align}
Note that sometimes, the derivative of $f_\pot$ with respect to the density is decomposed into a rearrangement term $\Sigma_R$ related to the explicit density dependence of the interaction, or Lagrangian, from the rest, see the subsection dedicated to the density-dependent relativistic mean-field model.

The kinetic pressure $p^*_{\kin,q}$ can also be expressed in terms of the Fermi-Dirac distribution $F_D$,
\begin{align}
F_D(k,\mu^*_{\kin,q},T,m^*_q) = \left[1+e^{\left((k^2+m^{*2}_q)^{1/2}-\mu^*_{\kin,q}\right)/T}\right]^{-1},
\end{align}
where $\mu^*_{\kin,q}$ is the chemical potential at finite $T$, defined as 
\begin{align}
\mu^*_{\kin,q} &= \frac{\partial \phi^*_{\kin,q}}{\partial n_q}\Bigg|_{T,n_{\bar{q}},\{\varphi\}} \, ,
\label{eq:effmass}
\end{align}
in which $q$ represents a particle of a given isospin-index, and ${\bar q}$ describes the other one.

The relativistic kinetic energy density and the kinetic pressure are defined as
\begin{align}
\epsilon^*_{\kin,q} &= \frac{\gamma}{2\pi^2}\int_0^{\infty}\hspace{-0.2cm}dk\,k^2(k^2 + 
m^{*2}_q)^{1/2}F_D(k,\mu^*_{\kin,q},T,m^*_q)\, ,
\label{eq:edenfg} \\
p^*_{\kin,q} &= \frac{\gamma}{6\pi^2}\int_0^{\infty}\hspace{-0.2cm}\frac{dk\,k^4}{(k^2 + 
{m^{*2}_q})^{1/2}}F_D(k,\mu^*_{\kin,q},T,m^*_q)\, ,
\label{eq:pressfg}
\end{align}
where $\gamma=2$ is the spin degeneracy for spin saturated systems. The nucleon entropy density is defined as,
\begin{equation}
\sigma_q = -\frac{\gamma}{2\pi^2}\int_0^{\infty}dk\,k^2
[F_D\mbox{ ln} F_D + (1-F_D)\mbox{ln}(1-F_D)]\, ,
\label{eq:entdenfg}
\end{equation}
with $\sigma=\sigma_n+\sigma_p$.

By fixing $\hbar=c=k_B=1$, momenta, masses, and temperatures are given in units of energy. 
For simplicity, we disregard here possible anti-particle contributions but they can be simply added to the formalism.

The neutrons and protons chemical potentials $\mu_q$ read
\begin{align}
\mu_q &= \frac{\partial \phi}{\partial n_q}\Bigg|_{T,n_{\bar{q}}} = \frac{\partial\phi^*_{\kin}}{\partial n_q}\Bigg|_{T,n_{\bar{q}},\{\varphi\}}
+ \frac{\partial\phi_{\pot}}{\partial n_q}\Bigg|_{T,n_{\bar{q}},\{\varphi\}}
\nonumber\\
&+ \sum_{i}\frac{\partial\phi}{\partial \varphi_{i}}\Bigg|_{T,n_q,n_{\bar q},\{\varphi_{{j\neq i}}\}}\frac{\partial \varphi_{i}}{\partial n_q}\Bigg|_{T,n_{\bar q}}
\nonumber\\
&= \mu^*_{\kin,q} + \mu_{{\corr},q} + \mu_{\pot,q}\, ,
\label{eq:muq}
\end{align}
with $\mu^*_{\kin,q}$ defined from Eq.~\eqref{eq:effmass} and
\begin{align}
&\mu_{{\corr},q} = \sum_{i}
\frac{\partial \varphi_{i}}{\partial n_q}\Bigg|_{T,n_{\bar q}} 
\frac{\partial \left( \phi^*_{\kin}+\phi_{\pot}\right)}{\partial \varphi_{i}}\Bigg|_{T,n_q,n_{\bar{q}},\{\varphi_{{j\neq i}}\}}
\, ,
\label{eq:mupcorr}\\
&\mu_{\pot,q} = \frac{\partial\phi_{\pot}}{\partial n_q}\Bigg|_{T,n_{\bar{q}},\{\varphi\}}\, .
\end{align}

In relativistic approaches, the scalar density is often introduced, since it arises naturally in the coupling between nucleons and scalar fields. It also contributes to the saturation mechanism, since vector and scalar fields interact with nucleons with different vertex defining different densities.
The scalar density for neutrons and protons is defined as
\begin{align}
n_{s,q} = \frac{\gamma m^*_q}{2\pi^2}\int_0^\infty \hspace{-0.2cm}\frac{dk\,k^2}{(k^2+m_q^{*2})^{1/2}}F_D(k,\mu^*_{\kin,q},T,m^*_q)\, ,
\label{eq:nsq}
\end{align}
and the isoscalar scalar density is $n_s=n_{s,n}+n_{s,p}$.
One could demonstrate that the scalar field $n_{s,q}$ can be expressed in terms of kinetic energy density, pressure and effective mass as follows:
\begin{align}
n_{s,q} = \frac{\epsilon^*_{\kin,q} - 3p^*_{\kin,q}}{m^*_q} \, .
\label{eq:e3p}
\end{align}
As shown in Eq.~\eqref{eq:e3p}, the scalar density can be determined from the thermodynamical quantities given by the JEL approximation. This is what we have done to obtain the equation of state within the JEL approximation shown in Figs.~\ref{fig:rmfnl}-\ref{fig:rmfdd}.

\section{Application to phenomenological models}
\label{sec:examples}

In this section we present applications of the aforementioned formalism to some of the most widely employed phenomenological models used to describe nuclear physics systems. 

\subsection{Skyrme model}

We start by considering Skyrme model~\citep{skyrme1,skyrme2,stone2007,dutra2012}, for which the energy density can be written as the sum of the rest mass and the internal energy densities as,
\begin{align}
\epsilon^{\sky} = \epsilon_\mass + \epsilon^{\sky}_{\intern}\, ,
\end{align}
with $\epsilon_\mass=\sum_q m_q n_q$ and the internal energy expressed as
\begin{align}
\epsilon^{\sky}_{\intern} = \sum_q \epsilon^{\sky *}_{\intern\kin,q} \,+ \epsilon^{\sky}_{\pot}\, ,
\end{align}
where
\begin{align}
\epsilon^{\sky *}_{\intern\kin,q} &= \frac{\gamma}{2\pi^2}\int_0^{\infty}\hspace{-0.2cm}dk\,k^2\frac{k^2}{2m^*_q}F_D(k,\mu^*_{\kin,q},T,m^*_q)\, ,
\label{eq:edenkinsky}
\end{align}
and
\begin{align}
\epsilon^{\sky}_{\pot}(n,\delta) &= \frac{1}{8}t_0n^2[2(x_0+2)-(2x_0+1)H_2]
\nonumber\\
&+ \frac{1}{48}t_{3}n^{\alpha+2}[2(x_{3}+2)-(2x_{3}+1)H_2]\, ,
\label{eq:edenpotsky}
\end{align}
with
\begin{align}
H_2 = \frac{1}{2}[(1-\delta)^2 + (1+\delta)^2]\, .
\end{align}

For Skyrme model, the fields are the effective masses, $\{\varphi\}=\{m^*_n, m^*_p\}$, which are defined in terms of $n$ and $\delta$ as, 
\begin{equation}
\frac{m^*_q(n,\delta)}{m}=\left[1+2m\left(C_0^\tau+\tau_3C_1^\tau \delta\right)n\right]^{-1} \, ,
\label{eq:effmasssky}
\end{equation}
where $m$ is the nucleon bare mass and $\tau_3=1$ for neutrons and $-1$ for protons. Here there are seven model parameters which are: $x_0$, $t_0$, $x_3$, $t_3$, $\alpha$, $C_0^\tau$, $C_1^\tau$.

According to~\cite{dutra2023finite}, the entropy density, $\sigma^{\sky}$, does not present any correction due to the in-medium effective mass $m^*_q$ in Skyrme model, $\sigma^{\sky}=\sigma^{\sky *}$, since the effective nucleon mass is independent on $T$, see Eq.~\eqref{eq:effmasssky}.
It is therefore possible to express the Helmholtz free energy~\eqref{eq:Helmholtz} as
\begin{align}
\phi^{\sky} &\equiv \epsilon^{\sky} - T\sigma^{\sky} = \epsilon^{\sky} - T\sigma^{\sky *} 
\end{align}
which gives
\begin{align}
\phi^{\sky} &\equiv \phi_\mass + \phi^{\sky *}_{\intern\kin} + \phi^{\sky}_{\pot}
\label{eq:freesky}
\end{align}
with
\begin{align}
&\phi_{\mass} = \epsilon_{\mass}\, , \\
&\phi^{\sky *}_{\intern\kin} = \sum_{q=n,p} \epsilon^{\sky *}_{\intern\kin,q} \,-\, T\sigma^{\sky *} \, , \\
&\phi^{\sky}_{\pot} = \epsilon^{\sky}_{\pot}\, ,
\label{eq:phipotsky}
\end{align}
where $\epsilon^{\sky *}_{\intern\kin,q}$ and $\sigma^{\sky *}$ are obtained directly from the analytical approximation of the Fermi integrals at fixed effective mass. The Helmholtz free energy $\phi^{\sky}$ is therefore directly obtained from the analytical expressions without in-medium correction.

For Skyrme model, the potential term is independent of the fields $\{\varphi\}$, 
see Eq.~\eqref{eq:edenpotsky}, which implies 
$\partial\epsilon^{\sky}_{\pot}/\partial m^*_q=0$, and 
using Eq.~\eqref{eq:phipotsky}, we obtain
$\partial\phi^{\sky}_{\pot}/\partial m^*_q=0$.
As a result, we obtain the following expression for the pressure:
\begin{align}
p^{\sky} = \sum_{q=n,p} \left(p^{\sky *}_{\kin,q}+  
p^{\sky}_{\corr,q} \right)
\,+\, p^{\sky}_{\pot} 
\end{align}
with the following contributions:
\begin{align}
&p^{\sky *}_{\kin,q} = \frac 2 3 \epsilon^{\sky *}_{\intern\kin,q} \, , \label{eq:sky:pkin}\\
&p^{\sky}_{\corr,q} = - \frac{3}{2}n\frac{p^{\sky *}_{\kin,q}}{m^*_q}\frac{\partial m^*_q}{\partial n}\Bigg| _{T,\delta}\label{eq:sky:pcorr}\\
&p^{\sky}_{\pot} = \frac{1}{8}t_0n^2[2(x_0+2)-(2x_0+1)H_2]
\nonumber\\
&\hspace{1cm}+ \frac{1}{48}t_{3}(\alpha+1)n^{\alpha+2}[2(x_{3}+2)-(2x_{3}+1)H_2]\, ,
\end{align}
where the correction term \eqref{eq:sky:pcorr} implying the derivative of $\phi$ with respect to the in-medium effective mass is obtained directly from Eq.~\eqref{eq:pcorr} using the relation
\begin{equation}
\frac{\partial \phi^{\sky *}_{\intern\kin}}{\partial m^*_q}\Bigg|_{T,n,\delta} = \frac{1}{m^*_q}(\epsilon^{\sky *}_{\intern\kin,q}-3p^{\sky *}_{\kin,q}) \, ,
\label{eq:pcorr-asym}
\end{equation}
and injecting the relation~\eqref{eq:sky:pkin}. 

In particular, Eq.~\eqref{eq:pcorr-asym} was derived in Ref.~\cite{dutra2023finite} by taking into account the analytical expressions furnished by the JEL approximation. We address the reader to this reference for more details on this calculation. Since $\epsilon_{\kin,q}^*=m_q n_q + \epsilon_{\intern\kin,q}^*$, one can use the relation~\eqref{eq:e3p} to express 
\begin{equation}
\frac{\partial \phi^{\sky *}_{\intern\kin}}{\partial m^*_q}\Bigg|_{T,n,\delta} = n_{s,q} - \frac{m_q}{m^*_q} n_q \, .
\end{equation}

Note that the pressure in Skyrme model contains a correction term $p_{\corr,q}^\sky$ due to the in-medium effective mass given by the Eq.~\eqref{eq:sky:pcorr}.

For the chemical potentials, we have
\begin{align}
\mu_q^{\sky} &= m_q + \mu^*_{\kin,q} + \mu_{\corr,q}^{\sky} + \mu_{\pot,q}^{\sky}
\end{align}
with $\mu^*_{\kin,q}$ defined from Eq.~\eqref{eq:effmass} and
\begin{align}
\mu_{\corr,q}^{\sky} &= - \frac{3}{2}\frac{p^{\sky *}_{\kin,n}}{m^*_n}\frac{\partial m^*_n}{\partial n_q}\Bigg| _{T,n_{\bar{q}}}
- \frac{3}{2}\frac{p^{\sky *}_{\kin,p}}{m^*_p}\frac{\partial m^*_p}{\partial n_q}\Bigg| _{T,n_{\bar{q}}}
\nonumber\\
\mu_{\pot,q}^{\sky} &= \frac{t_0}{4}n\left\{2(x_0+2)-(2x_0+1)[H_2\pm(1\mp\delta)\delta]\right\}
\nonumber\\
&\hspace{1cm}+\frac{(\alpha+2)}{48}t_3n^{\alpha+1}\Bigg\{2(x_3+2)
\nonumber\\
&\hspace{1cm}-(2x_3+1)\left[H_2\pm\frac{2(1\mp\delta)\delta}{\alpha+ 2}\right]\Bigg\}\, ,
\label{eq:muqsky}
\end{align}
with upper (lower) signs for neutrons (protons).

\subsection{Nonlinear relativistic mean-field model}

The energy density of nonlinear relativistic mean-field (RMF) models~\citep{boguta77,baoanli2008,dutra2014,batista} with fixed coupling constants, denoted here as a nonlinear (NL) model, can be expressed as
\begin{align}
\epsilon^{\NL} = \sum_{q=n,p}\epsilon^{*}_{\kin,q} \,+\, \epsilon^{\NL}_{\pot}\, ,
\end{align}
with the kinetic energy density defined in Eq.~\eqref{eq:edenfg} and the potential term expressed as
\begin{align}
\epsilon^{\NL}_{\pot} &= \frac{1}{2}m^2_\sigma\sigma^2_0 
+ \frac{A}{3}\sigma_0^3 + \frac{B}{4}\sigma_0^4 - \frac{1}{2}m^2_\omega\omega_0^2 
- \frac{C}{4}(g_\omega^2\omega_0^2)^2 \nonumber\\
&- \frac{1}{2}m^2_\rho \rho_{0(3)}^2
+g_\omega\omega_0n-\frac{g_\rho}{2}\rho_{0(3)}n_3 
\nonumber \\
&+ \frac{1}{2}m^2_\delta\delta^2_{(3)} - g_\sigma g_\omega^2\sigma_0\omega_0^2\left(\alpha_1+\frac{1}{2}{\alpha'_1}g_\sigma\sigma_0\right) \nonumber\\
&- g_\sigma g_\rho^2\sigma_0\rho_{0(3)}^2 
\left(\alpha_2+\frac{1}{2}{\alpha'_2} g_\sigma\sigma_0\right) - \frac{1}{2}{\alpha'_3}g_\omega^2 g_\rho^2\omega_0^2\rho_{0(3)}^2,
\label{eq:denl}
\end{align}
where $n_3=n_n-n_p=\delta n$. 
Here $\sigma_0$, $\delta_{(3)}$, $\omega_0$ and $\rho_{0(3)}$ are the mean-field reductions of the meson fields with masses $m_\sigma$, $m_\delta$, $m_\omega$, and $m_\rho$. The coupling constants of the model are given by $g_\sigma$, $g_\omega$, $g_\rho$, $g_\delta$, $A$, $B$, $C$, $\alpha_1$, $\alpha'_1$, $\alpha_2$, $\alpha'_2$, and $\alpha'_3$. The effective nucleon mass is given in terms of the scalar fields $\sigma_0$ and $\delta_{(3)}$, namely,
\begin{align}
m^*_q = m^*_q(\sigma_0,\delta_{(3)}) = m_q - g_\sigma\sigma_0 + \tau_3g_\delta\delta_{(3)}.
\label{eq:effmassnl}
\end{align}

The field equations for the fields $\sigma_0$ and $\delta_{(3)}$, deduced from the Euler-Lagrange equations, are given by
\begin{align}
&m^2_\sigma\sigma_0 = g_\sigma(n_{s,n}+n_{s,p}) - A\sigma_0^2 - B\sigma_0^3 
\nonumber \\
&+g_\sigma g_\omega^2\omega_0^2(\alpha_1+\alpha'_1g_\sigma\sigma_0)
+g_\sigma g_\rho^2\rho_{0(3)}^2(\alpha_2+\alpha'_2g_\sigma\sigma_0),
\end{align}
and
\begin{eqnarray}
m_\delta^2\delta_{(3)} &= -g_\delta(n_{s,n}-n_{s,p}), 
\end{eqnarray}
which shows that these fields are modified by the medium, mostly from the scalar densities, see Eq.~\eqref{eq:nsq}. 
Similar relations could be obtained for the other fields.

Since the effective mass can be expressed in terms of the meson fields, see Eq.~\eqref{eq:effmassnl}, and the fields are in-medium quantities, the field contribution to the Helmholtz free energy in the relativistic mean-field model can be developed as: $\{\varphi\}=\{\sigma_0,\delta_{(3)},\omega_0,\rho_{0(3)}\}$.


The entropy density is given by
\begin{align}
\sigma^{\NL} &= 
-\frac{\partial\phi^{\NL}}{\partial T}\Bigg|_{n,\delta} 
= -\frac{\partial\phi^{\NL}_{\kin}}{\partial T}\Bigg|_{n,\delta,\{\varphi\}} 
- \frac{\partial\phi^{\NL}_{\pot}}{\partial T}\Bigg|_{n,\delta,\{\varphi\}} \nonumber \\
&\hspace{3cm} -\sum_i\frac{\partial \phi^{\NL}}{\partial\varphi_i}\Bigg|_{n,\delta,T,\varphi_{{j\neq i}}}\frac{\partial\varphi_i}{\partial T}\Bigg|_{n,\delta} \, .
\label{entrmf}
\end{align}
We remark that i) the potential term in the RMF model does not depend explicitly on $T$, so the second term in Eq.~\eqref{entrmf} vanishes, and ii) the equilibrium relation~\citep{walecka_book} in the CE leads to
\begin{align}
\frac{\partial\phi^{\NL}}{\partial\varphi_i}\Bigg|_{n,\delta,T,\varphi_{{j\neq i}}} = 0 \, ,
\label{gibbs}
\end{align}
which then cancels the last term in
Eq.~\eqref{entrmf}. We thus obtain that
\begin{align}
\sigma^{\NL} = -\frac{\partial\phi^{\NL}_{\kin}}{\partial T}\Bigg|_{n,\delta,\{\varphi\}} = \sigma^{\NL*} \, ,
\end{align}
which means that the entropy density can be directly obtained from the JEL approximation, with no in-medium correction.
The Helmholtz free energy can then be expressed as
\begin{align}
\phi^{\NL} &\equiv \epsilon^{\NL} - T\sigma^{\NL} = \epsilon^{\NL} - T\sigma^{\NL *} \equiv \phi^{\NL *}_{\kin} + \phi^{\NL}_{\pot}
\label{eq:freeNL}
\end{align}
with
\begin{align}
\phi^{\NL *}_{\kin} &= \sum_{q=n,p}\epsilon^{\NL *}_{\kin,q} \,- T\sigma^{\NL *}\, , \\ 
\phi^{\NL}_{\pot} &= \epsilon^{\NL}_{\pot}\, .
\end{align}
The pressure is therefore obtained as
\begin{align}
p^{\NL} &=\sum_{q=p,n} p^*_{\kin,q} \,+ p^{\NL}_{\corr} + p^{\NL}_{\pot}\, ,
\label{eq:pressnl}
\end{align}
with $p^*_{\kin,q}$ given by the relation~\eqref{eq:pressfg} and can be calculated from the JEL approximation with in-medium effective mass, as shown in Ref.~\cite{dutra2023finite}, and 
\begin{align}
p^{\NL}_{\corr} &= n\sum_i \frac{\partial\varphi_i}{\partial n}\Bigg|_{\delta,T} \frac{\partial\phi^{\NL}}{\partial\varphi_i}\Bigg|_{n,\delta,T,\varphi_{{j\neq i}}} = 0\, , \\
p^{\NL}_{\pot} &= n\frac{\partial\epsilon^{\NL}_{\pot}}{\partial n}\Bigg|_{T,\delta,\{\varphi\}} - \epsilon^{\NL}_{\pot} \, ,
\end{align}
where we have used the equilibrium condition~\eqref{gibbs} to show that $p^{\NL}_{\corr}=0$. In other words, there is no correction term to the pressure induced by the in-medium effective mass, at variance with Skyrme model.

The final expression for $p^{\NL}$ is
\begin{align}
p^{\NL} &= \sum_{q=n,p} p^{*}_{\kin,q} \, - \frac{1}{2}m^2_\sigma\sigma_0^2 -\frac{A}{3}\sigma_0^3 - \frac{B}{4}\sigma_0^4    
\nonumber\\
&+ \frac{1}{2}m^2_\omega\omega_0^2 + \frac{C}{4}(g_\omega^2\omega_0^2)^2 + \frac{1}{2}m^2_\rho \rho_{0(3)}^2
- \frac{1}{2}m^2_\delta\delta^2_{(3)}
\nonumber \\
&+ g_\sigma g_\omega^2\sigma_0\omega_0^2\left(\alpha_1+\frac{1}{2}{\alpha'_1}g_\sigma\sigma_0\right) + \frac{1}{2}{\alpha'_3}g_\omega^2 g_\rho^2\omega_0^2\rho_{0(3)}^2
\nonumber \\
&+ g_\sigma g_\rho^2\sigma_0\rho_{0(3)}^2 \left(\alpha_2+\frac{1}{2}{\alpha'_2} g_\sigma\sigma_0\right)\, .
\label{eq:pressnlfinal}
\end{align}
Note that all the terms linear in the density $n$ do not contribute to the pressure. The pressure~\eqref{eq:pressnlfinal} coincides with the expression obtained directly from the momentum-energy tensor~\citep{baoanli2008,dutra2014}.

Finally, the chemical potentials of the model are 
\begin{align}
\mu^{\NL}_q &= \frac{\partial\phi^{\NL}}{\partial n_q}\Bigg|_{T,n_{\bar q}} 
\nonumber \\
&= \mu^*_{\kin,q} + 
\sum_i\frac{\partial\varphi_i}{\partial n_q}\Bigg|_{T,n_{\bar{q}}}\frac{\partial \phi^{\NL}}{\partial\varphi_i}\Bigg|_{n_q,n_{\bar{q}},T,\varphi_{{j\neq i}}}
\nonumber\\
&+\frac{\partial\epsilon^{\NL}_{\pot}}{\partial n_q}\Bigg|_{T,n_{\bar{q}},\{\varphi\}}\, .
\end{align}
Once again, Eq~\eqref{gibbs} leads to $\mu_{\corr,q}=0$ and 
\begin{align}
\mu^{\NL}_q &= \mu^*_{\kin,q} + \frac{\partial\epsilon^{\NL}_{\pot}}{\partial n_q}\Bigg|_{T,n_{\bar{q}},\{\varphi\}}
\nonumber\\
&= \mu^*_{\kin,q} + g_\omega\omega_0 \mp \frac{g_\rho}{2}\rho_{0(3)}\, ,
\end{align}
with $-$ ($+$) for neutrons (protons). Note that similarly to the pressure, there is no correction to the chemical potential induced by the in-medium effective fields.

\subsection{Density-dependent relativistic mean-field model}

Another widely used nucleonic model is the one in which the couplings are density-dependent functions~\citep{typel1999,gogelein,baoanli2008,dutra2014}, namely,
\begin{align}
\epsilon^{\DD} = \sum_{q=n,p}\epsilon^{*}_{\kin,q} \,+ \epsilon^{\DD}_{\pot}\, ,
\end{align}
with 
\begin{align}
\epsilon^{\DD}_{\pot} &= \frac{1}{2}m^2_\sigma\sigma_0^2 - \frac{1}{2}m^2_\omega\omega_0^2 - \frac{1}{2}m^2_\rho \rho_{0(3)}^2
+ \frac{1}{2}m^2_\delta\delta^2_{(3)} 
\nonumber \\
&+ \Gamma_\omega(n)\omega_0n - \frac{\Gamma_\rho(n)}{2}\rho_{0(3)}n_3,
\label{eq:dedd}
\end{align}
where the functions $\Gamma_j$ ($j=\sigma,\omega,\rho,\delta$) are given by polynomial or fractional forms in terms of the density~\citep{typel1999,gogelein,sidney}. 
The in-medium effective masses for the neutrons and the protons are defined as
\begin{align}
m^*_n &= m - \Gamma_\sigma(n)\sigma_0 + \Gamma_\delta(n)\delta_{(3)}\, ,
\label{eq:effmassddn} \\
m^*_p &= m - \Gamma_\sigma(n)\sigma_0 - \Gamma_\delta(n)\delta_{(3)}\, ,
\label{eq:effmassddp}
\end{align}
As in the NL model, there are four fields in the theory: $\{\varphi\}=\{\sigma_0, \delta_{(3)}, \omega_0,\rho_{0(3)}\}$. 

A similar to the nonlinear RMF case analysis can be performed here, and we obtain 
\begin{align}
&\frac{\partial\epsilon^{\DD}_{\pot}}{\partial n}\Bigg|_{T,\delta,\{\varphi\}} = 
\Gamma_\omega\omega_0 - \frac{\Gamma_\rho}{2}\rho_{0(3)}\delta + \Sigma_R(n)
\end{align}
with the rearrangement term $\Sigma_R$ defined as
\begin{align}
\Sigma_R(n) = \Gamma'_\omega\omega_0 n - \frac{\Gamma'_\rho}{2}\rho_{0(3)}n_3 - \frac{m_\sigma^2\sigma_0^2\Gamma'_\sigma}{\Gamma_\sigma}
-\frac{m_\delta^2\delta_{(3)}^2\Gamma'_\delta}{\Gamma_\delta}\, ,
\end{align}
where $\Gamma'_j\equiv d\Gamma_j/dn$. 

Finally, we obtain the following expression for the pressure
\begin{align}
p^{\DD} &= \sum_{q=n,p} p^{*}_{\kin,q}  \,+ n\Sigma_R (n) - \frac{1}{2}m^2_\sigma\sigma_0^2 + \frac{1}{2}m^2_\omega\omega_0^2  
\nonumber\\
&\hspace{2cm}+ \frac{1}{2}m^2_\rho \rho_{0(3)}^2 - \frac{1}{2}m^2_\delta\delta^2_{(3)}\, .
\label{eq:pressddfinal}
\end{align}

For the chemical potentials, we have
\begin{align}
\mu_q^{\DD} = \mu^*_{\kin,q} + \Sigma_R (n) + \Gamma_\omega\omega_0 \mp \frac{\Gamma_\rho}{2}\rho_{0(3)}\, ,
\end{align}
with $-$ ($+$) for neutrons (protons).

\subsection{Metamodel}
The nonrelativistic version of the metamodel developed in Refs.~\cite{mm1,mm2,mm-up1,mm-up2} considers the energy density given by the sum of kinetic and potential parts with the latter expressed in terms of an expansion in the parameter $x=(n-n_\sat)/(3n_\sat)$, where $n_\sat$ is the saturation density. At finite temperature, it reads
\begin{align}
\epsilon^\mm(n,T,\delta) = m\,n + \epsilon^{\mm *}_{\intern\kin,p} + \epsilon^{\mm *}_{\intern\kin,n} + \epsilon_{\rm pot}^{\mm}(n,\delta),
\end{align}
with the nonrelativistic kinetic energy density of protons and neutrons given as in the Skyrme model, i.e., by the expression shown in Eq.~(\ref{eq:edenkinsky}), with the following nucleon effective mass
\begin{align}
\frac{m}{m^*_q(n,\delta)} = 1 + (\kappa_\sat + \tau_3\kappa_\sym\delta)\frac{n}{n_\sat}.
\end{align}
The potential part of the model is written as
\begin{align}
\epsilon_{\rm pot}^{\mm}(n,\delta) = n\sum_{j=0}^N\frac{1}{j!}(v_{\sat,j} + v_{\sym,j}\delta^2)x^j u_j(x,\delta),
\end{align}
where
\begin{align}
 u_j(x,\delta) = 1 - (-3x)^{N+1-j}e^{-\zeta(\delta)(3x+1)},
\end{align}
and $\zeta(\delta)=b_\sat+b_\sym\delta^2$. Here we take $N=4$ and use $b_\sat=6.9$, $b_\sym=0$~\cite{mm1,mm2}. The coefficients $v_{\sat,j}$, $v_{\sym,j}$, and the parameters $\kappa_\sat$, $\kappa_\sym$ are given in terms of the nuclear empirical parameters, whose values can be given in the respective ranges~\citep{mm1}
\begin{align}
n_\sat &= (0.155 \pm 0.005)\,\rm fm^{-3} \, ,
\\
E_\sat &= (-15.8 \pm 0.3)\,\rm MeV \, ,
\\
\frac{m^*_\sat}{m} &= \frac{m_q^*(n_\sat,0)}{m}= 0.75 \pm 0.1 \, ,
\\
K_\sat &= (230\pm 20)\,\rm MeV \, ,
\\
Q_\sat &= (300\pm 400)\,\rm MeV \, ,
\\
Z_\sat &= (-500\pm 1000)\,\rm MeV,
\\
\frac{\Delta m^*_\sat}{m} &= \frac{m^*_n(n_\sat,1)}{m}-\frac{m^*_p(n_\sat,1)}{m}
= 0.1 \pm 0.1\, ,
\\
E_\sym &= (32 \pm 2)\,\rm MeV \, ,
\\
L_\sym &= (60 \pm 15)\,\rm MeV \, ,
\\
K_\sym &= (-100\pm 100)\,\rm MeV \, ,
\\
Q_\sym &= (0\pm 400)\,\rm MeV \, ,
\\
Z_\sym &= (-500\pm 1000)\,\rm MeV\, .
\end{align}
For the calculations performed in this paper, we adopted the central values of each interval presented above. 

As in the case of the Skyrme model, we verify that the nucleon entropy density of the metamodel is also given by Eq.~\eqref{eq:entdenfg}, since its effective mass is a temperature independent quantity. Therefore, we have
\begin{align}
\phi^{\mm}(n,T,\delta) &= \epsilon^{\mm *}_{\kin,p} + \epsilon^{\mm *}_{\kin,n} - T(\sigma^{\mm *}_p + \sigma^{\mm *}_n) 
\nonumber\\
&+ \epsilon_{\rm pot}^{\mm}(n,\delta).
\label{eq:freemm}
\end{align}
Therefore, for usual calculations, the pressure and chemical potentials of the model can be numerically calculated from this main thermodynamical quantity, namely,
\begin{align}
p^\mm(n,T,\delta) &= n^2 \frac{\partial (\phi^\mm/n)}{\partial n}\Bigg|_{T,\delta},
\label{eq:presmm_num}
\\
\mu_q^\mm &= \frac{\partial \phi^\mm}{\partial n_q}\Bigg|_{T,n_{\bar q}}.
\label{eq:chempotmm_num}
\end{align}
However, the procedure developed in Sec.~\ref{sec:formalism} is a useful tool in order to derive analytical expressions in replacement of the numerical ones. For the pressure of the metamodel, for instance, we obtain
\begin{align}
&p^{\mm}(n,T,\delta) = \frac{2}{3}(\epsilon^{\mm *}_{\kin,p} + \epsilon^{\mm *}_{\kin,n}) 
\nonumber\\
&- n\left( \frac{\epsilon^{\mm *}_{\kin,p}}{m^*_p}\frac{\partial m^*_p}{\partial n}\Bigg| _{T,\delta} + \frac{\epsilon^{\mm *}_{\kin,n}}{m^*_n}\frac{\partial m^*_n}{\partial n}\Bigg| _{T,\delta} \right) + p^{\mm}_{\pot}(n,\delta),
\label{eq:presmm}
\end{align}
with 
\begin{align}
&p_{\rm pot}^{\mm}(n,\delta) = 
\nonumber\\
&=\frac{n^2}{3n_\sat}\left[\sum_{i=0}^{N-1}\frac{1}{i!}(v_{\sat,i+1} + v_{\sym,i+1}\delta^2)x^i u_{i+1}(x,\delta)\right.
\nonumber\\
&+ \left. \sum_{j=0}^{N}\frac{1}{j!}(v_{\sat,j} + v_{\sym,j}\delta^2)x^j w_j(x,\delta) \right],
\end{align}
and
\begin{align}
&w_j(x,\delta) = \frac{\partial u_j}{\partial x}
\nonumber\\
&= 3(-3x)^{N-j}e^{-\zeta(\delta)(3x+1)}[N+1-j - 3x\zeta(\delta)].
\end{align}
In addition, the chemical potential of the nucleon $q$ reads
\begin{align}
\mu_q^{\mm} &= m + \mu^*_{\kin,q} 
- \frac{\epsilon^{\mm *}_{\kin,n}}{m^*_n}\frac{\partial m^*_n}{\partial n_q}\Bigg| _{T,n_{\bar{q}}}
- \frac{\epsilon^{\mm *}_{\kin,p}}{m^*_p}\frac{\partial m^*_p}{\partial n_q}\Bigg| _{T,n_{\bar{q}}}
\nonumber\\
&+ \mu_{\pot,q}^{\mm},
\end{align}
where
\begin{align}
\mu_{\pot,q}^{\mm} &= \frac{1}{n}(\epsilon_\pot^\mm + p_\pot^\mm) + \frac{\partial\epsilon_\pot^\mm}{\partial\delta}\frac{\partial\delta}{n_q},
\end{align}
with
\begin{align}
&\frac{\partial\epsilon_\pot^\mm}{\partial\delta} = 
\nonumber\\
&=n\sum_{j=0}^{N}\frac{x^j}{j!}\Big[\delta v_{\sym,j} u_j(x,\delta) + (v_{\sat,j} + v_{\sym,j}\delta^2)v_j(x,\delta)\Big],
\end{align}
and
\begin{align}
v_j(x,\delta) = \frac{\partial u_j}{\partial\delta} 
= 2(3x+1)b_\sym\delta\,[1 - u_j(x,\delta)].
\end{align}
The derivative $\partial\delta/\partial n_q$ is equal to $-2n_n/n^2$~($2n_p/n^2$) for $n_q=n_p$~($n_q=n_n$).

These analytical expressions are important, for example, in order to reduce time consumption in complex computational methods, such as the Bayesian analysis, implemented in general along with Markov chain Monte-Carlo, in which a huge number of configurations are performed in each run.

\subsection{Numerical implementation}

In order to show how the aforementioned formalism is applied, we compute in this section all the previous thermodynamical quantities for all phenomenological models at different temperatures. For this purpose, it is needed to choose a suitable treatment for the Fermi integrals. There are several of them in the literature as the reader can see, for instance, in~\cite{eggleton1973,Antia-1993,pols,JEL-1996-1020,Aparicio_1998,mohankumar-2005,natarajan-2001,mamedov-2012,fukushima-2014,khvorostukhin-2015,gil-2023} and references therein. Here we use the one proposed in~\cite{JEL-1996-1020}, named as JEL approximation and also used in our previous study~\citep{dutra2023finite}, in which the Fermi integrals are described in terms of analytical functions. We compare such an approach with a typical numerical calculation using the Gauss-Legendre method with 600 Gauss points. We display the results for two specific parameterizations of the Skyrme and RMF models, namely, SLy4~\citep{sly4} and BSR1~\citep{bsr1}, for the density dependent model \mbox{DD-ME2}~\cite{ddme2}, and for the metamodel used here. They were shown to be consistent with experimental data regarding ground state binding energies, charge radii, and giant monopole resonances of some finite nuclei, as well as in good agreement with stellar matter properties, according to the findings of~\cite{prc2023}.

\begin{figure*}[!htb]
\centering
\includegraphics[scale=0.85]{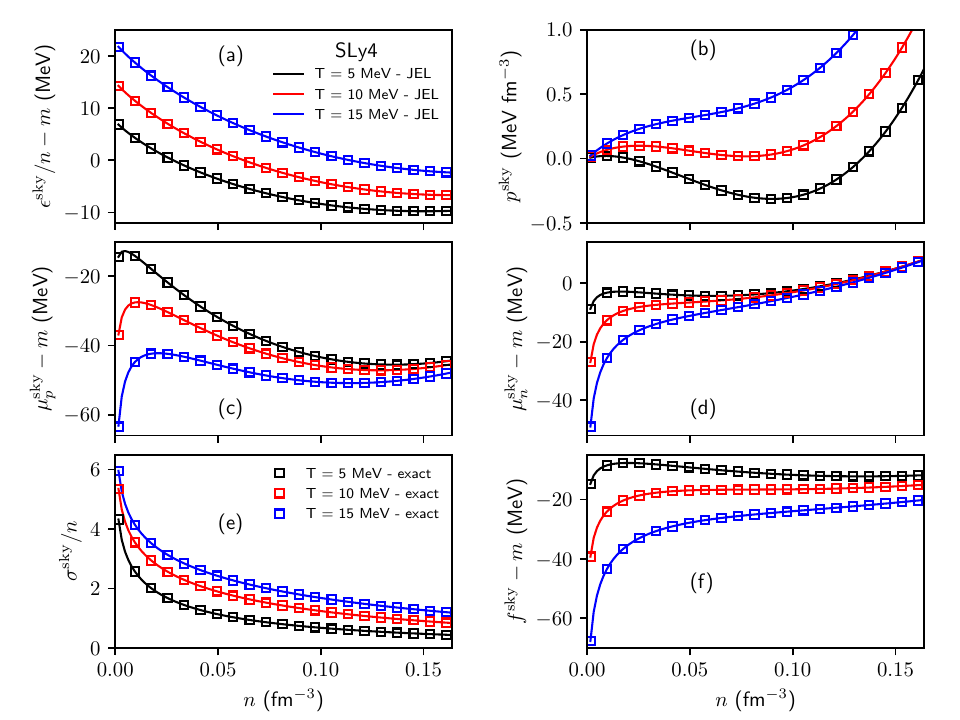}
\caption{Thermodynamic quantities of nonrelativistic SLy4 model~\citep{sly4}. Exact calculation (squares) and JEL approximation (full lines), for different temperatures and $\delta=0.4$: (a) energy per particle $\epsilon^{\rm sky}/n - m$, (b) pressure $p^{\sky}$, (c) proton chemical potential $\mu^{\sky}_p - m$, (d) neutron chemical potential $\mu^{\sky}_n - m$, (e) entropy per particle $\sigma^{\sky}/n$, and (f) Helmholtz free energy per particle $f^{\sky} - m =\phi^{\sky}/n - m$.} 
\label{fig:sky}
\end{figure*}

The results for the SLy4 parametrization concerning energy per particle, pressure, chemical potentials, entropy per particle, and Helmholtz free energy per particle are depicted in Fig.~\ref{fig:sky}.

Note the very good overlap between the JEL approximation and the exact calculation. In order to quantify better this agreement, we calculate the residual difference defined as
\begin{equation}
\xi_X = \sqrt{\frac{1}{N}\sum_{i=1}^{N}\left(\frac{X_{\jel, i} - X_{\mbox{\tiny{exact}},i}}{X_{\mbox{\tiny{exact}},i}}\right)^2}, 
\label{eq:chi}
\end{equation}
where $N$ is the number of points, $X_{\jel, i}$ is the thermodynamical function calculated through the JEL approximation, and $X_{\mbox{\tiny{exact}},i}$ is the same quantity obtained by performing the exact calculation (numerical integration). The numbers are presented in the first three lines of Table~\ref{tab:chi}.
\begin{table*}[tb]
\centering
\tabcolsep=0.4cm
\def\arraystretch{1.2}
\caption{Residual difference defined in Eq.~\eqref{eq:chi} between the exact calculations and the JEL approximation of the thermodynamical quantities of the respective parametrizations of the phenomenological models presented in Figs.~\ref{fig:sky}-\ref{fig:mm}.}
\begin{tabular}{lccccccc}
\hline\hline
Model & $T$   & $\xi_{\epsilon/n}$ & $\xi_{p}$  & $\xi_{\mu_p}$ & $\xi_{\mu_n}$ & $\xi_{\sigma/n}$ & $\xi_{f}$\\ 
& (MeV) & ($10^{-4}$)  & ($10^{-4}$) & ($10^{-5}$) &  ($10^{-5}$) & ($10^{-5}$) &($10^{-5}$)\\ 
\hline    
\multirow{3}{*}{SLy4~\citep{sly4}} & $~~5$ & $1.34$  & $4.83$ & $66.9$ & $25.1$ & $1.84$ & $2.48$\\
                      & $10$  & $0.54$  & $3.58$ & $52.3$ & $14.3$ & $1.01$ & $0.63$\\
                      & $15$  & $0.28$  & $3.46$ & $45.6$ & $10.8$ & $0.89$ & $0.21$\\
\cline{2-8}
\multirow{3}{*}{BSR1~\citep{bsr1}} & $~~5$ & $16.5$  & $8.92$ & $0.02$ & $0.03$ & $1.10$ & $91.1$\\
                      & $10$  & $9.76$  & $0.43$ & $0.02$ & $0.03$ & $1.18$ & $47.1$\\
                      & $15$  & $5.46$  & $0.15$ & $0.02$ & $0.03$ & $0.99$ & $19.9$\\
\cline{2-8}
\multirow{3}{*}{DD-ME2~\citep{ddme2}} & $~~5$ & $0.16$  & $0.61$ & $0.22$ & $3.07$ & $10.5$ & $0.59$\\
                      & $10$  & $0.67$  & $0.92$ & $1.40$ & $11.7$ & $3.89$ & $1.01$\\
                      & $15$  & $1.91$  & $0.85$ & $0.28$ & $18.9$ & $2.94$ & $0.50$\\
\cline{2-8}
\multirow{3}{*}{Metamodel} & $~~5$ & $8.60$ & $7.12$ & $61.2$ & $18.9$ & $1.43$ & $1.85$\\
                      & $10$  & $0.79$  & $3.69$ & $49.7$ & $12.7$ & $0.97$ & $0.49$\\
                      & $15$  & $0.32$  & $3.52$ & $43.8$ & $10.1$ & $0.86$ & $0.20$\\                      
\hline
\end{tabular}
\label{tab:chi}
\end{table*}

\begin{figure*}[!htb]
\centering
\includegraphics[scale=0.85]{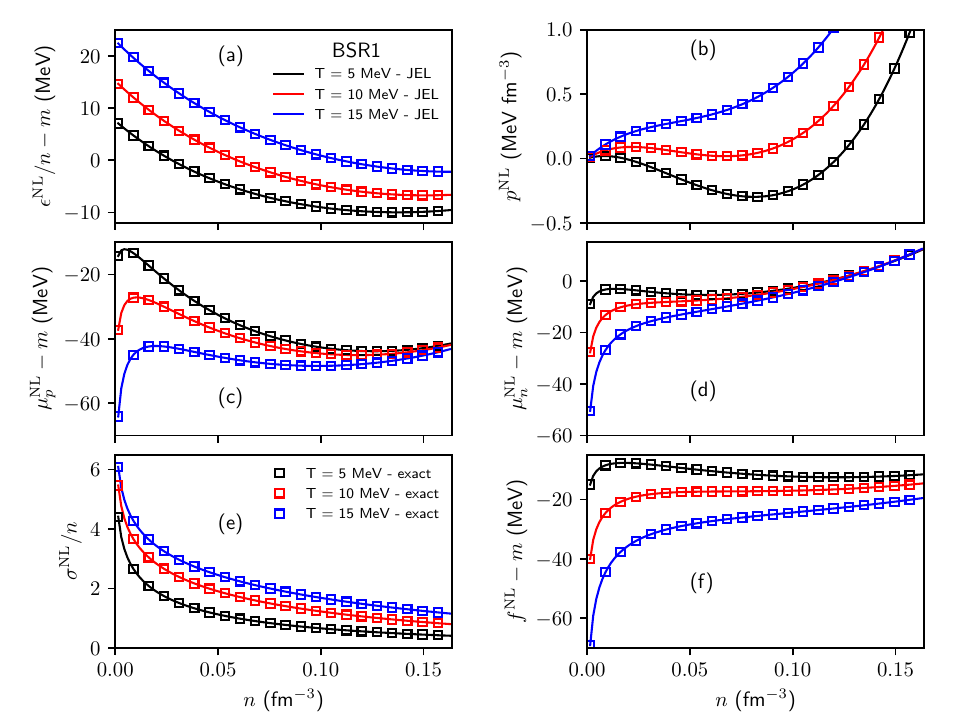}
\caption{The same as Fig.~\ref{fig:sky} for the nonlinear relativistic model BSR1~\citep{bsr1}.} 
\label{fig:rmfnl}
\end{figure*}

\begin{figure*}[!htb]
\centering
\includegraphics[scale=0.85]{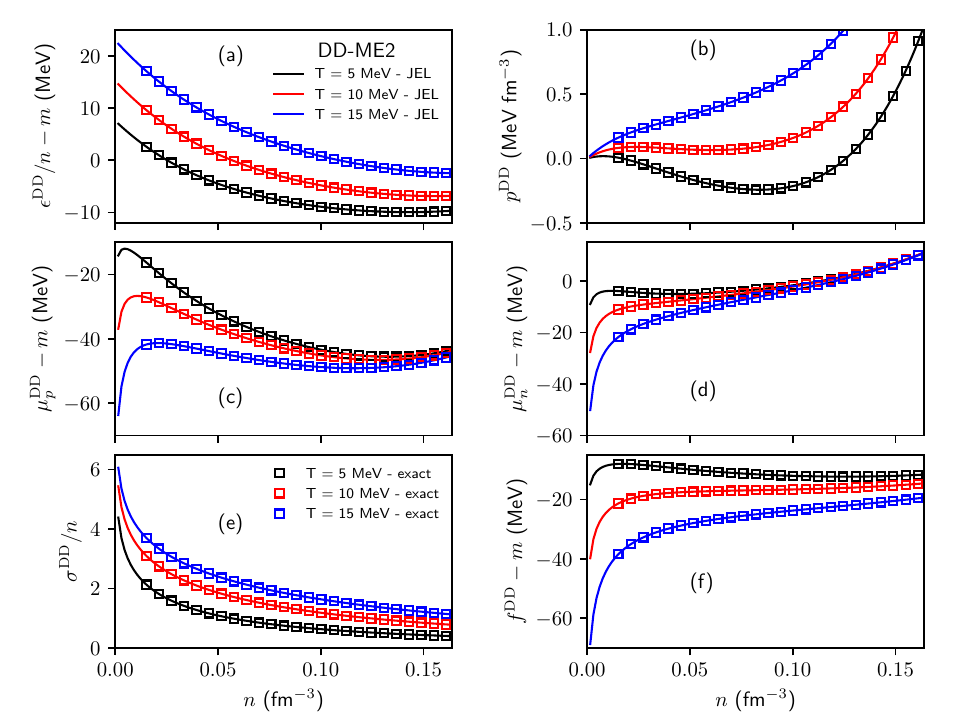}
\caption{The same as Fig.~\ref{fig:sky} for the density-dependent relativistic model \mbox{DD-ME2}~\citep{ddme2}.}
\label{fig:rmfdd}
\end{figure*}

\begin{figure*}[!htb]
\centering
\includegraphics[scale=0.85]{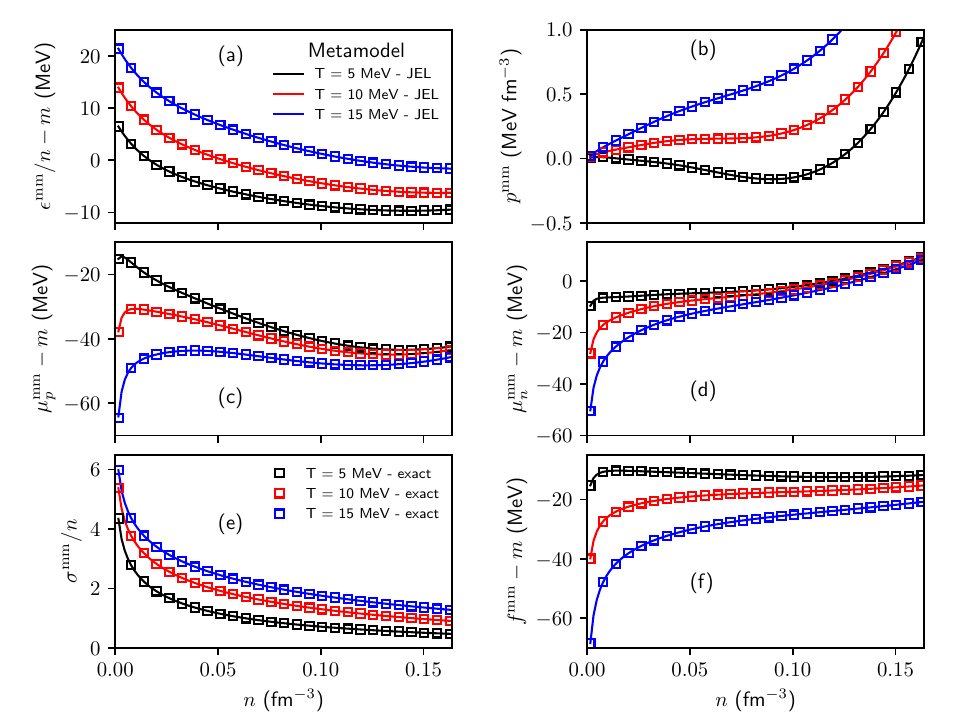}
\caption{The same as Fig.~\ref{fig:sky} for the metamodel.}
\label{fig:mm}
\end{figure*}

The same thermodynamical quantities predicted by BSR1~\citep{bsr1} nonlinear and \mbox{DD-ME2}~\citep{ddme2} density-dependent relativistic models, as well as the metamodel, are displayed respectively in Figs.~\ref{fig:rmfnl}-\ref{fig:mm}. 
Note here also that the JEL approximation provides a very accurate approximation of the exact calculation. The JEL approximation can therefore be safely used as an alternative to the numerical integration even for the relativistic case. As in the previous case, we quantify the comparison between the exact calculation and the JEL approximation through Eq.~\eqref{eq:chi}. The numbers are shown in Table~\ref{tab:chi}. 

As a last remark, we emphasize the efficiency of the JEL approximation in comparison with the numerical integration (600 Gauss points) used in this work. For the Skyrme model we find that the JEL approximation is about 15 times faster than the numerical calculation. This factor is changed to about 30 in the case of the nonlinear relativistic model.

\section{Conclusions}
\label{sec:conclusions}

In this paper, we have performed an improvement of the recent study presented in~\cite{dutra2023finite} where a systematic analysis of FG at finite temperature with in-medium effect taken into account by effective masses. More specifically, we now consider a generic nucleonic model with the respective Helmholtz free energy density depending on the effective fields. 

We have provided the generalized thermodynamical quantities for this case and have shown examples of three widely used models, namely, Skyrme, nonlinear, and density-dependent relativistic mean-field models, as well as the metamodel. We have also evaluated the equations of state as functions of the density, and for different values of temperature, by numerically solving the Fermi integrals and we have compared it with the analytical formulation proposed in~\cite{JEL-1996-1020}, and we have also used in~\cite{dutra2023finite}. It was shown that considering the proper in-medium corrections to the thermodynamical quantities, generally defined as the equation of state, one could safely employ analytical approximates of the Fermi integrals at finite temperature, such as for instance the JEL approximation, to compute the properties of nuclear matter at finite temperature. 

\begin{acknowledgements}
This work is a part of the project INCT-FNA proc. No. 464898/2014-5. It is also supported by Conselho Nacional de Desenvolvimento Cient\'ifico e Tecnol\'ogico (CNPq) under Grants No. 312410/2020-4, 307255/2023-9 (O.~L.), and No. 308528/2021-2 (M.~D.). O.~L. and M.~D. also acknowledge CNPq under Grant No. 401565/2023-8 (Universal). O.~L. is also supported by FAPESP under Grant No. 2022/03575-3 (BPE). This study was financed in part by the Coordena\c{c}\~ao de Aperfei\c{c}oamento de Pessoal de N\'ivel Superior - Brazil (CAPES) - Finance Code 001 - Project number 88887.687718/2022-00 (M.~D.).
J.M. is supported by the CNRS-IN2P3 MAC masterproject, and benefits from the LABEX Lyon Institute of Origins (ANR-10-LABX-0066) of the \textsl{Universit\'e de Lyon}.
\end{acknowledgements}

\bibliographystyle{apsrev4-2}
\bibliography{references-revised}

\end{document}